\pdfoutput=1

\documentclass[aps,pra,twocolumn,superscriptaddress]{revtex4}

\usepackage{color}
\usepackage{units}
\usepackage{graphicx}

\begin{document}

\title{A pump-probe study of the formation of rubidium molecules by the ultrafast photoassociation of ultracold atoms}

\author{David J. McCabe}
\email{d.mccabe1@physics.ox.ac.uk}
\author{Duncan G. England}
\author{Hugo E. L. Martay}
\author{Melissa E. Friedman}
\author{Jovana Petrovic}
\affiliation{Clarendon Laboratory, Department of Physics, University of Oxford, Oxford, OX1 3PU, United Kingdom}
\author{Emiliya Dimova}
\affiliation{Institute of Solid State Physics, 72 Tzarigradsko Chaussee, 1784 Sofia, Bulgaria}
\author{B\'{e}atrice Chatel}
\affiliation{ CNRS-Universit\'e de Toulouse,UPS, Laboratoire Collisions, Agr\'egats R\'eactivit\'e,
IRSAMC, F-31062 Toulouse, France}
\author{Ian A. Walmsley}
\affiliation{Clarendon Laboratory, Department of Physics, University of Oxford, Oxford, OX1 3PU, United Kingdom}

\date{\today}

\begin{abstract}
An experimental pump-probe study of the photoassociative creation of translationally ultracold rubidium molecules is presented together with numerical simulations of the process. The formation of loosely bound excited-state dimers is observed as a first step towards a fully coherent pump-dump approach to the stabilization of Rb$_2$ into its lowest ground vibrational states. The population that contributes to the pump-probe process is characterized and found to be distinct from a background population of pre-associated molecules.

Published in Phys.\ Rev.\ A 80, 033404 (2009) (DOI: 10.1103/PhysRevA.80.033404)
\end{abstract}


\maketitle

\section{\label{Intro} Introduction}

The successful development of atomic laser-cooling techniques has facilitated the study of atomic collisions in the ultracold regime and the macroscopic quantum degeneracy of the Bose--Einstein condensate \cite{Cornell2002,Ketterle2002}. This progress has inspired interest in the generation of ultracold molecules in prescribed low-energy internal states, opening prospects from ultrahigh precision spectroscopy to the coherent control of chemical reactions and new techniques for quantum computing \cite{DeMille2002,Tesch2002}. Atomic cooling techniques, however, cannot be generalized even to simple diatomic molecules due to the lack of availability of a closed-loop cooling cycle within the rich molecular internal energy level structure; instead, alternative approaches must be pursued.

Various alternative direct cooling techniques have been developed that stabilize molecules efficiently into the lowest vibrational levels, but are unable to cool their translational motion significantly below the millikelvin regime \cite{Weinstein1998,Bethlem2000}. In order to reach microkelvin or nanokelvin temperatures, it is necessary instead to associate ultracold atoms via approaches such as photoassociation \cite{Thorsheim1987,Fioretti1998}, the manipulation of Feshbach resonances \cite{Greiner2003,Zwierlein2003}, or three-body collisions \cite{Jochim2003}. These processes, however, favour the formation of ultracold but vibrationally excited molecules.

Recently, an incoherent optical cycling scheme was demonstrated that accumulates population in the Cs$_2$ singlet ground vibrational state via spontaneous decay \cite{Viteau2008}. Precise knowledge of the system spectroscopy is not required; however it relies upon the serendipitous availability within Cs$_2$ of a specific continuous-wave (c.w.) pump-decay photoassociative pathway that pre-populates deeply bound vibrational states. Such a pathway is not, in general, available in other dimers, and an optical repumping scheme would require many more spontaneous decay cycles (and hence greater heating of the centre-of-mass motion) before significant enhancement of the ground vibrational state population is obtained. In certain heteronuclear systems, high-power, c.w.\ photoassociation has been used to produce an incoherent mixture that includes $v=0$ ground-state singlet dimers \cite{Sage2005,Deiglmayr2008}. In another approach, stimulated Raman adiabatic passage (STIRAP) has been used to transfer molecules to the lowest vibrational ground state in Rb$_2$ triplets \cite{Lang2008} and Cs$_2$ singlets \cite{Danzl2009,Mark2009}, as well as heteronuclear KRb molecules to both triplet and singlet $v=0$ states \cite{Ni2008}. The transfer is coherent but requires full spectroscopic knowledge of the system.

%Coherent control \cite{Warren1993,Rabitz2000} --- the application of \new{phase-}tailored optical fields to \new{manipulate} the course of a chemical reaction --- presents a complimentary strategy to those outlined above.
A complementary quantum control strategy to those outlined above is the coherent control \cite{Warren1993,Rabitz2000} of the photoassociation process using broadband tailored optical fields. The broad coherent bandwidth of a Ti:sapphire ultrafast pulse is well matched to typical diatomic molecule potential binding energies, while the carrier frequency is appropriate for ground- to excited-state transitions. An ultracold scattering event may therefore be steered towards a target state through a variety of vibronic pathways, with phase and amplitude shaping of the excitation pulse offering control over the system Hamiltonian via the dipole interaction. Furthermore, closed-loop control may be applied to the choice of pulse shape in order to identify an empirical optimum. The generality of this approach is thus limited merely by the frequency and bandwidth of the laser system rather than any requirement for precise spectroscopic knowledge.

The use of ultrafast `pump' and `dump' pulses would permit the fully coherent transfer of population from free atom pairs to deeply bound vibrational states \cite{Koch2006a}. An excited-state molecular wavepacket is formed by photoassociation, and evolves freely before being optically de-excited at a time that yields optimal Franck--Condon overlap with the target state. %This coherent superposition of excited states thus offers a pathway to deeply bound product molecules despite the poor Franck--Condon factors of the individual vibrational states. \new{Through exploitation of a time--non-stationary coherent superposition in the excited state, this wavepacket approach thus offers the advantage that the population transfer may be effected with favourable Franck--Condon factors for both `pump' and `dump' stages without relying upon the overlap between individual wavefunctions of specific vibrational states}.
Through the exploitation of a non-stationary coherent superposition in the excited state, this wavepacket approach thus offers the advantage of favourable Franck--Condon factors for both `pump' and `dump' stages without relying upon the overlap between individual wavefunctions of specific vibrational states. Previous ultrafast `pump-decay' photoassociation experiments have revealed a coherent quenching effect on pre-formed triplet molecules \cite{Brown2006,Staanum2006}; a pump-dump approach may enable the coherent enhancement of bound molecular population before the onset of spontaneous decay.

Learning about the dynamics present in the excited state after photoassociation is therefore an important step towards efficient stabilization into the ground state. A useful tool for resolving ultrafast molecular dynamics is the pump-probe experiment \cite{Machholm1994,Fatemi2001}. A pump pulse generates an excitation which evolves during a controllable time delay before being detected by a probe pulse through a reaction-coordinate--selective mechanism such as ionization or bond fragmentation. Recently, a pump-probe study of the short-delay dynamics of rubidium atoms and molecules after photoassociation revealed coherent transient oscillations during the photoassociation pulse \cite{Salzmann2008}. In this paper we report on pump-probe experiments investigating the coherent control of bound excited-state dimers over longer timescales as the first step towards the development of a pump-dump scheme to produce deeply bound ground-state Rb$_2$ singlet molecules. We demonstrate the production of bound excited-state molecules and compare the experimental observations to numerical simulations. We characterize the initial state of a background molecular population within the cold atom cloud and identify their association mechanism. %In order to permit improved comparison between theory and experiment, as well as to gain further insight into this reaction process, we infer the origin of the population addressed by the pump pulse and contrast the contributions of bound and unbound initial populations to the pump-probe signal. 
We show evidence that the pump pulse addresses free atom pairs with internuclear separations of \unit[30-60]{a$_0$}, rather than the closer range background molecules. This experimental determination of the initial conditions permits improved comparison between theory and experiment, offering further insight into the reaction process.

\section{Methods and apparatus}
\label{sec:method}

\subsection{Overview}
\label{sec:overview}

  \begin{figure}
    \centering
    \includegraphics[width=\columnwidth]{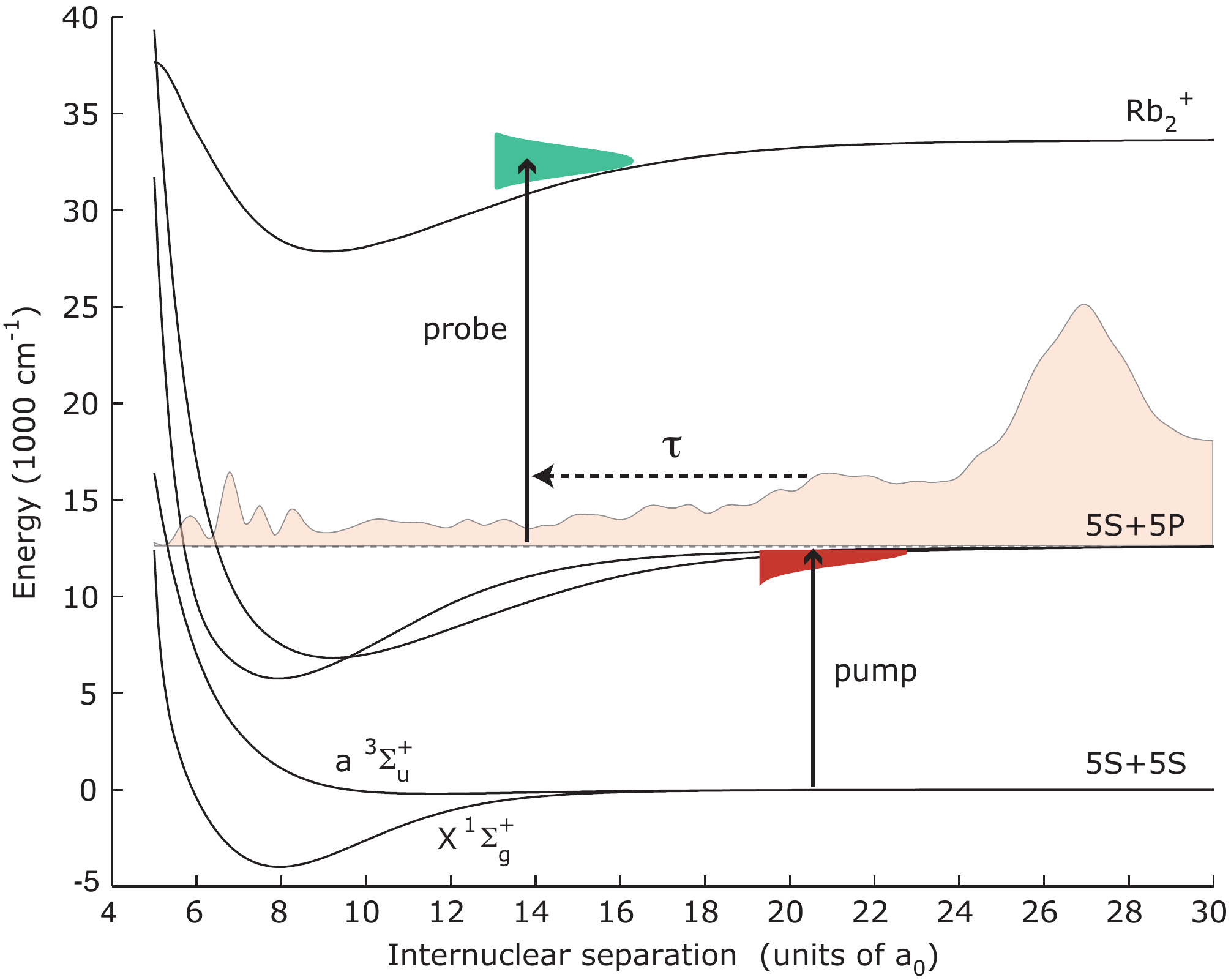}
    \caption{(Colour online). A schematic of the pump-probe experiment. A femtosecond pump pulse photoassociates atom pairs from the ground state, forming a wavepacket spanning a range of vibrational states within the \mbox{5S + 5P} potential manifold. A typical simulated population distribution is shown (shaded). After a variable time delay $\tau$, a probe pulse ionizes the excited-state population in preparation for detection via TOF mass spectrometry.} 
    \label{fig:schematic}
  \end{figure}

\begin{figure}
    \centering
    \includegraphics[width=\columnwidth]{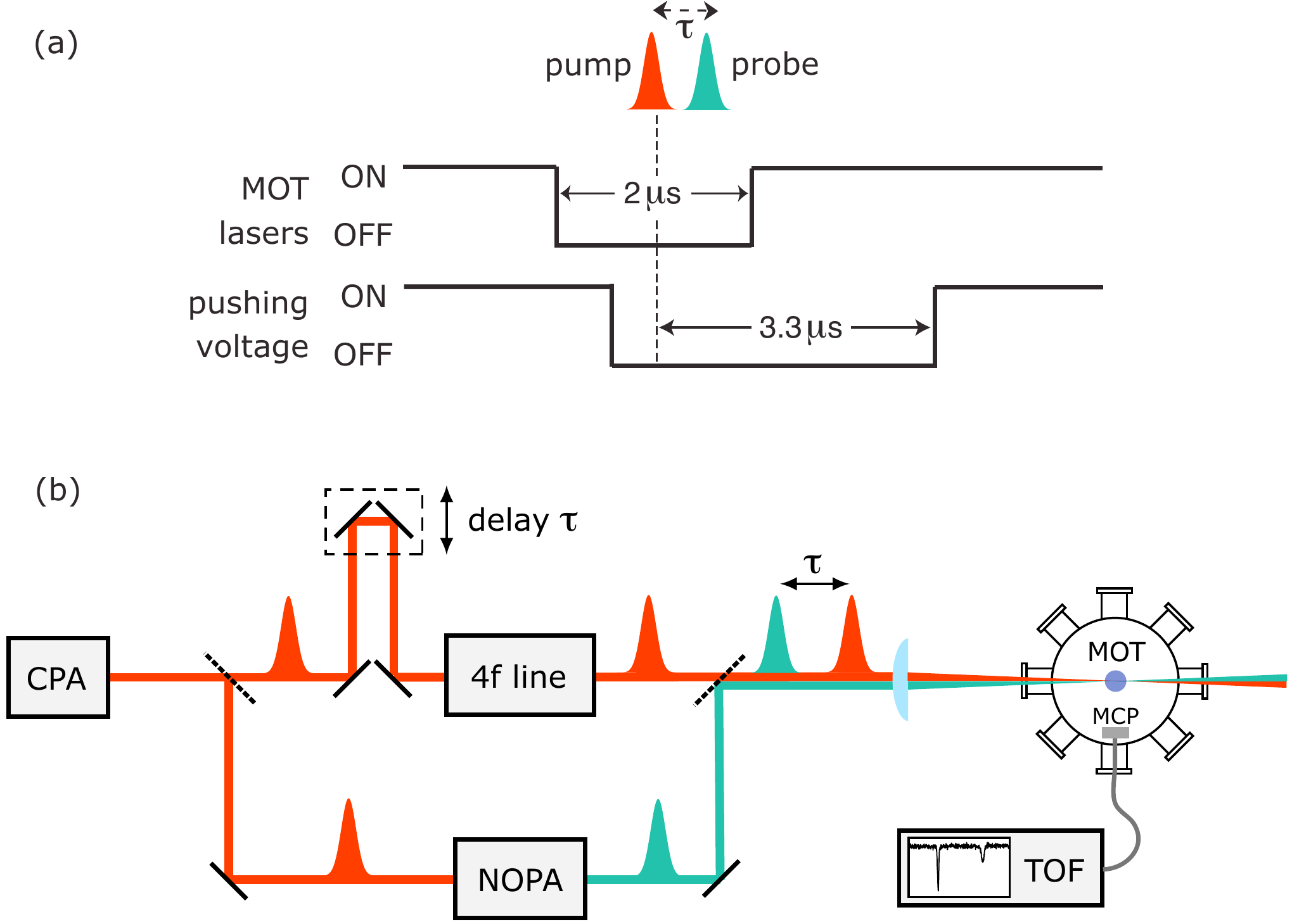}
    \caption{(Colour online). (a) A schematic of the experimental timing. The MOT lasers are shut off for a \unit[2]{$\mu$s} window spanning the arrival of the ultrafast pulses. The TOF pushing electrode is switched to a positive voltage \unit[3.3]{$\mu$s} after ionization. (b) Experimental layout. Pulses from a CPA laser pass through a beamsplitter, with the majority of the pulse energy pumping the NOPA to form the probe pulses whilst the remainder is spectrally filtered in a 4f line and acts as the pump. The pulses are combined with a variable delay $\tau$ and focussed onto the MOT. Ionized rubidium atoms and molecules are accelerated towards an MCP detector and distinguished via TOF mass spectrometry.}
    \label{fig:timing}
  \end{figure}

A schematic of the pump-probe experiment is given in Fig.\ \ref{fig:schematic}. A broadband pump pulse photoassociates ultracold ground-state 5S atom pairs within a rubidium magneto-optical trap (MOT), forming long-range molecular wavepackets within the \mbox{5S + 5P} potential manifold. The excited dimer is allowed to evolve before being ionized by a broadband probe pulse. In order to study the excited-state dynamics over the oscillation timescales of long-range molecules, this pump-probe delay is varied over up to \unit[250]{ps}. Atomic ions are also produced via nonresonant multi-photon processes. The nascent atomic and molecular ions are detected and distinguished via time-of-flight (TOF) mass spectrometry. In order to ensure that the atom pairs commence the experiment in their ground state, the MOT lasers (which, like the pump pulse, operate on the 5S $\rightarrow$ 5P transition) are shuttered for a \unit[2]{$\mu$s} window spanning the arrival of the pump and probe pulses (see Fig.\ \ref{fig:timing}(a)).
 
\subsection{Magneto-optical trap}
\label{sec:MOT}

The choice of rubidium for this study was motivated by the suitability of its atomic and molecular spectroscopy to Ti:sapphire wavelengths, as well as the extent to which its collisional properties and photoassociation spectroscopy have been studied and catalogued. It therefore provides a convenient test bed for the control scenarios, though the concepts are expected to be of much broader utility.

$^{85}$Rb atoms were cooled and trapped in a MOT generated in an ultrahigh vacuum stainless steel chamber with multiple windows for optical access. Three orthogonal pairs of counter-propagating beams of opposite circular polarizations intersected at the minimum of a quadrupole magnetic field generated by a pair of anti-Helmholtz current coils. Stray magnetic fields in the chamber were nulled using three orthogonal pairs of compensation coils. Rb was loaded into the MOT from a background vapour maintained by the application of a constant current to a series of standard commercial getters sources. 

The master trapping light was generated by a home-built external-cavity diode laser (ECDL). A pick-off of the trapping light was passed through a rubidium vapour cell, and a feedback loop was used to lock its frequency to a feature in the resultant Doppler-free saturated absorption spectrum. The master laser was in turn injected into a higher power, free-running slave diode laser. The output of the slave laser was detuned \unit[15]{MHz} to the red of the $\textrm{F}=3 \rightarrow \textrm{F}'=4$ trapping transition using an acousto-optic modulator (AOM) before passing through a single-mode polarization-maintaining fibre to the MOT. A total of \unit[30]{mW} trapping light was split equally over three retro-reflected beams of \unit[4.0]{mm} half-maximum diameter in intensity. A second ECDL was similarly locked to the $\textrm{F}=2 \rightarrow \textrm{F}'=3$ transition to repump population that accumulated in the dark $\textrm{F}=2$ lower hyperfine state, with a power of \unit[3]{mW} at the MOT. The AOMs were additionally employed as fast optical shutters for the trapping and repumping light.

The total fluorescence and density profile of the MOT cloud were monitored by a photodiode and pair of CCD cameras respectively. $4\times10^6$ atoms were trapped in the MOT with a maximum density of \unit[$1.1\times10^9$]{cm$^{-3}$} and a half-maximum diameter of \unit[0.8]{mm}. The MOT temperature was measured by the release-and-recapture method \cite{Lett1988} to be \unit[110]{$\mu$}K. 

\subsection{Ultrafast system}
\label{sec:ultrafast}

  \begin{figure}
    \centering
    \includegraphics[width=0.9\columnwidth]{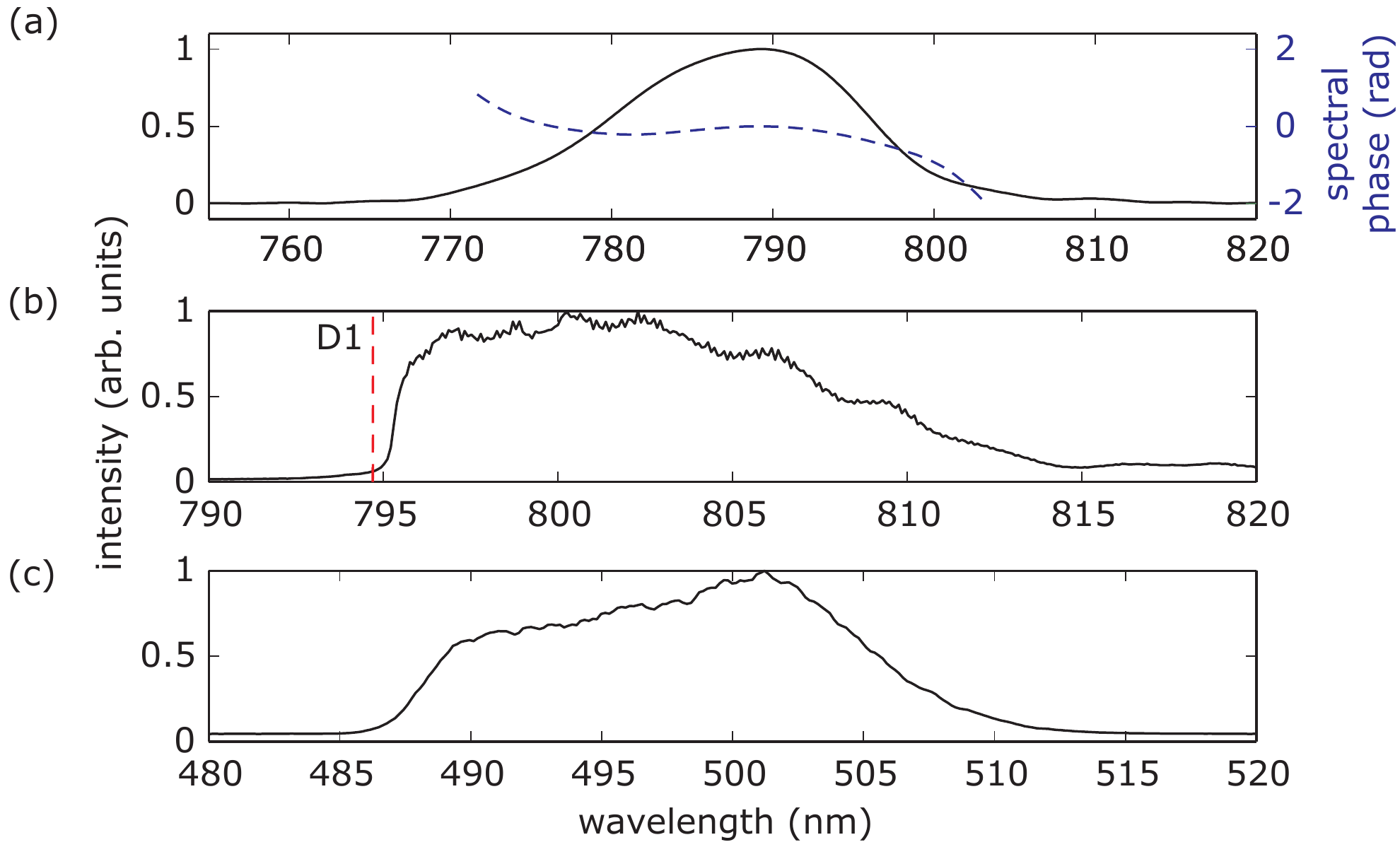}
    \caption{(Colour online). (a) Spectral intensity and phase of the CPA pulse used to generate the pump and probe (\unit[53]{fs} FWHM). (b) Typical pump spectrum with cut at Rb D1 line.(\unit[85]{fs} FWHM). (c) Typical NOPA probe spectrum, filtered at \unit[480]{nm} (\unit[390]{fs} FWHM).}
    \label{fig:pulses}
  \end{figure}

The ultrafast pulses for this experiment were derived from a regenerative chirped-pulse amplified Ti:sapphire laser (CPA) \cite{Backus1998} producing \unit[250]{$\mu$J}, \unit[800]{nm} pulses with a \unit[20]{nm} FWHM at a repetition rate of \unit[2]{kHz}. Pulse characterization using spectral phase interferometry for direct electric field reconstruction (SPIDER) \cite{Iaconis1998} revealed near--transform-limited pulses of \unit[53]{fs} FWHM (see Fig.\ \ref{fig:pulses}(a)). The CPA pulses were split into two components: a few microjoules were used to generate the pump pulse while the remainder pumped a non-collinear optical parametric amplifier (NOPA) which acted as the probe \cite{Cerullo2003}.

The pump pulse was spectrally filtered by a translatable razor blade in the Fourier plane of a zero-dispersion 4f line (\unit[1200]{lines/mm} diffraction gratings separated by \unit[75]{cm}) with a resolution of \unit[0.1]{nm} (as determined by the beam waist in the Fourier plane and the 4f line dispersion). The position of the spectral cut was set to eliminate excitation of unbound excited-state population above the \mbox{5S + 5P$_{1/2}$} `D1 line' atomic asymptote (see Fig.\ \ref{fig:pulses}(b)). Interferometric autocorrelation indicated pump pulses of \unit[85]{fs} FWHM duration. The NOPA spectrum was tuned to \unit[500]{nm} (\unit[20]{nm} FWHM) to correspond to the transitions between the \mbox{5S + 5P$_{1/2}$} and molecular ion states (see Fig.\ \ref{fig:pulses}(c)).  A long-pass filter with a \unit[480]{nm} cut-off prevented the resonant ionization of atoms. Intensity autocorrelation indicated probe pulses of \unit[390]{fs} FWHM duration. A typical pulse energy was \unit[4]{$\mu$J} with a fluctuation of less than \unit[5]{\%}.

The pump pulse was passed through a variable delay line, combined with the probe using a dichroic mirror (see Fig.\ \ref{fig:timing}(b)) and focussed onto the MOT.

\subsection{Detection scheme}
\label{sec:detection}

The efficiency of the photoassociation process is limited by the density of the MOT; hence the Rb$_2^+$ ionization rate is exceeded by Rb$^+$ ionization despite the off-resonant nature of the latter process. The challenge is therefore to resolve the molecular ions from the atomic ions with a sensitive detection mechanism. This was achieved using TOF mass spectrometry with a multi-channel plate (MCP) detector. The ions were accelerated towards the MCP by a gated pushing electrode situated opposite the MCP inside the vacuum chamber. The electrode was switched from \unit[0]{V} to a positive voltage \unit[3.3]{$\mu$s} after the arrival of the pump pulse. This provided for an expansion of the ion cloud that reduced the impact of saturation within the central MCP ion channels by distributing the ions over a greater detector surface area \cite{Kraft2007}.

The molecular ions arrived at the MCP within a time window narrower than \unit[100]{ns}. The sensitivity of the detection was changed by varying the pushing electrode voltage from \unit[100]{V} to \unit[1100]{V}: a high voltage ensured sensitive detection of  Rb$_2^+$ whilst a low voltage ensured a linear response to  Rb$^+$ without MCP saturation. For each pump-probe delay measurement the Rb$^+$ or Rb$_2^+$ TOF signal was averaged 1000 times by a digital oscilloscope and integrated numerically. This ensured that the duration of the single-point averaging was greater than that of the characteristic NOPA power fluctuations. The signal-to-noise ratio of the pump-probe signal was further improved without compromising immunity to long-term drifts by averaging several successive pump-probe scans.

\subsection{Bound initial state characterization}
\label{sec:REMPI}

\begin{figure}
  \centering
  \includegraphics[width=\columnwidth]{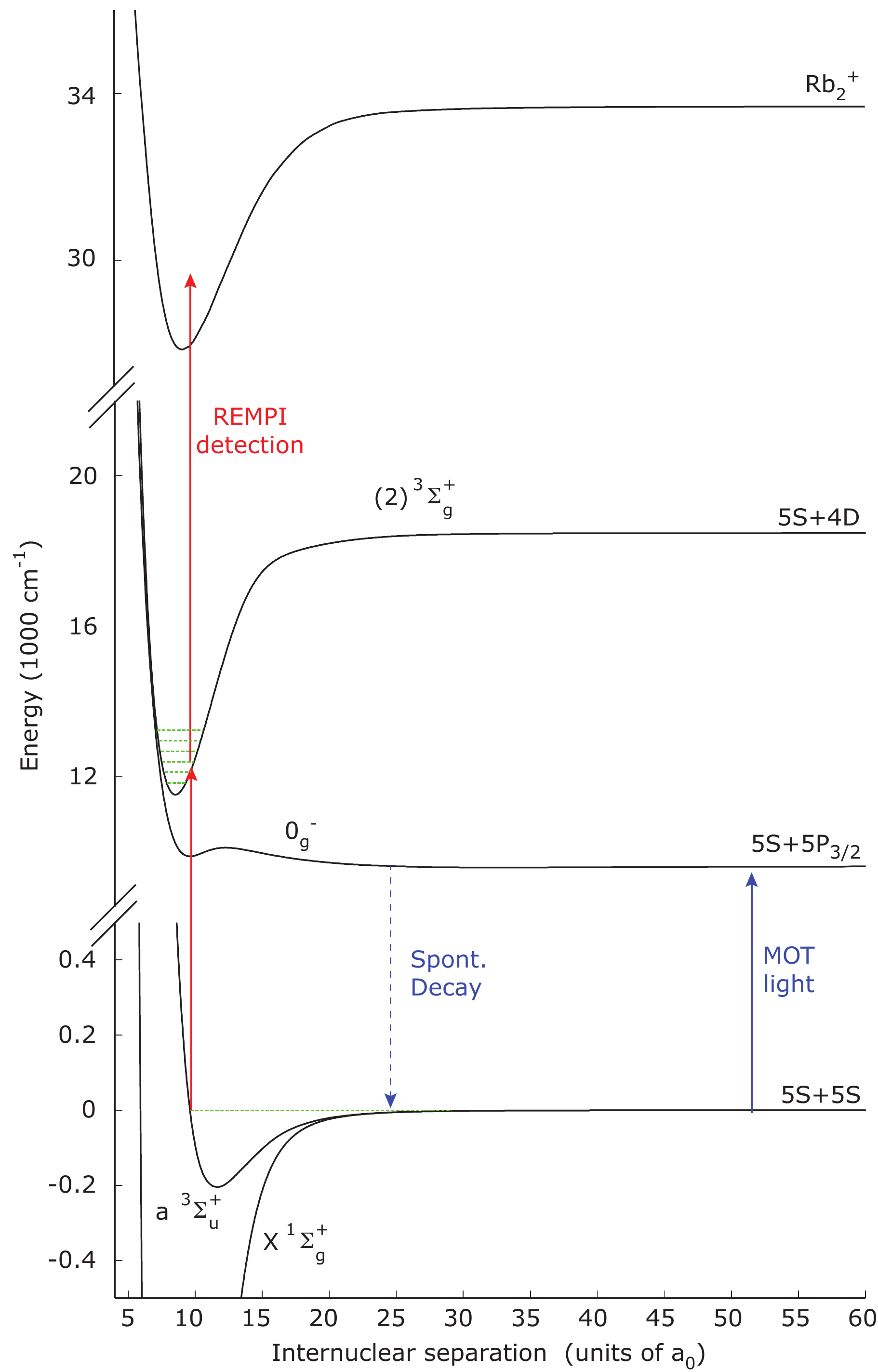}
  \caption{(Colour online). A schematic of the initial-state characterization experiment. The MOT trapping light photoassociates free atom pairs to loosely bound states within the \mbox{5S + 5P}$_{3/2}$ $0_{\textrm{g}}^-$ potential, from where they spontaneously decay to the $a^3\Sigma_u^+$ ground triplet potential. These `background' ground triplet molecules may be detected using a two-photon resonant ionization pathway via deeply bound states within the $(2)^3\Sigma_g^+$ potential. The high-lying populated states within the $a^3\Sigma_u^+$ ground state are more closely spaced than the deeply bound intermediates within the $(2)^3\Sigma_g^+$ potential, allowing the ground-state population distribution to be deduced from the resultant spectrum.}
  \label{fig:init_schematic}
\end{figure}

The intuitive picture of this pump-probe experiment concerns the photoassociation of a scattering pair of unbound atoms. There are, however, two mechanisms that produce a `background' population of loosely bound ground-state molecules in an $^{85}$Rb MOT in the absence of dedicated photoassociation light: three-body recombination \cite{Takekoshi1999}, and photoassociation by the MOT trapping light \cite{Fioretti1998}. Experimental studies in $^{85}$Rb have concluded that the former mechanism dominates at low trapping light intensities \cite{Gabbanini2000} but that the latter plays a more prominent role at higher intensities \cite{Caires2005}. 

It is possible that these preformed bound ground-state molecules play an important role in the pump-probe dynamics despite their paucity in comparison to unbound atom pairs, since the low atomic density within the MOT implies a low probability for two neighbouring atoms to interact. A precise characterization of the initial state of the interacting atom pairs is therefore essential for meaningful subsequent numerical simulation of the pump-probe process. The formation mechanism effected by the trapping light involves photoassociation predominantly to the \mbox{5S + 5P$_{3/2}$ $0_{\textrm{g}}^-$} potential followed by spontaneous decay to high-lying states within the $a^3\Sigma_u^+$ ground triplet potential (see Fig.\ \ref{fig:init_schematic}). This photoassociative mechanism is expected to dominate three-body recombination within our setup, as our experimental parameters more closely match those of Ref.\ \cite{Caires2005}. A comparison of the measured ground-state occupancy to the calculated population distribution resulting from this decay channel enables this hypothesis to be confirmed. 

The initial distribution of bound ground-state triplet molecules was studied via resonant multi-photon ionization (REMPI). A two-photon process preferentially ionized the Rb$_2$ via a resonant intermediate potential, while atomic ionization followed an off-resonant pathway and required an extra photon. A common REMPI pathway for detection of ground triplet Rb$_2$ employs loosely bound vibrational states within the $(2)^3\Pi_g$ potential as an intermediate \cite{Gabbanini2000}. Since the vibrational level spacings of these initial and intermediate states are comparable, it is not easy to unravel the initial distribution from a spectroscopic measurement via this pathway. Instead, this study uses an alternative REMPI route that passes through more deeply bound levels within the $(2)^3\Sigma_g^+$ potential (see Fig.\ \ref{fig:init_schematic}) \cite{Lozeille2006}. The resultant spectroscopic scan enables distinction between the large energy spacing of the intermediate states and the close spacing of the ground-state triplet Rb$_2$ whose occupancy it is desired to study.

An estimation of the population distribution of molecules photoassociated by the MOT lasers was made by considering the Franck--Condon overlap of high-lying $0_{\textrm{g}}^-$ states with each $a^3\Sigma_u^+$ vibrational state. The expected resultant spectrum for REMPI ionization via the $(2)^3\Sigma_g^+$ was calculated using the excited-state vibrational level energies given in Ref.\ \cite{Lozeille2006}, together with the corresponding transition Franck--Condon factors. A realistic bandwidth for the REMPI laser was assumed.

The light for the REMPI pathway outlined above was generated with a dye laser (using Pyridine 1 dye) that was pumped by the second harmonic of a Q-switched Nd:YAG laser. Pulses of \unit[10]{ns} duration and up to \unit[400]{$\mu$J} energy were produced at a repetition rate of \unit[50]{Hz}. The pulse wavelength was tunable between \unit[685]{nm} and \unit[705]{nm}, and was calibrated with reference to two-photon atomic transitions to the 4p$^6$6d configuration.

\section{Experimental results}
\label{sec:results}

\subsection{Bound initial state study}
\label{sec:initial-state}

\begin{figure}
  \centering
  \includegraphics[width=\columnwidth]{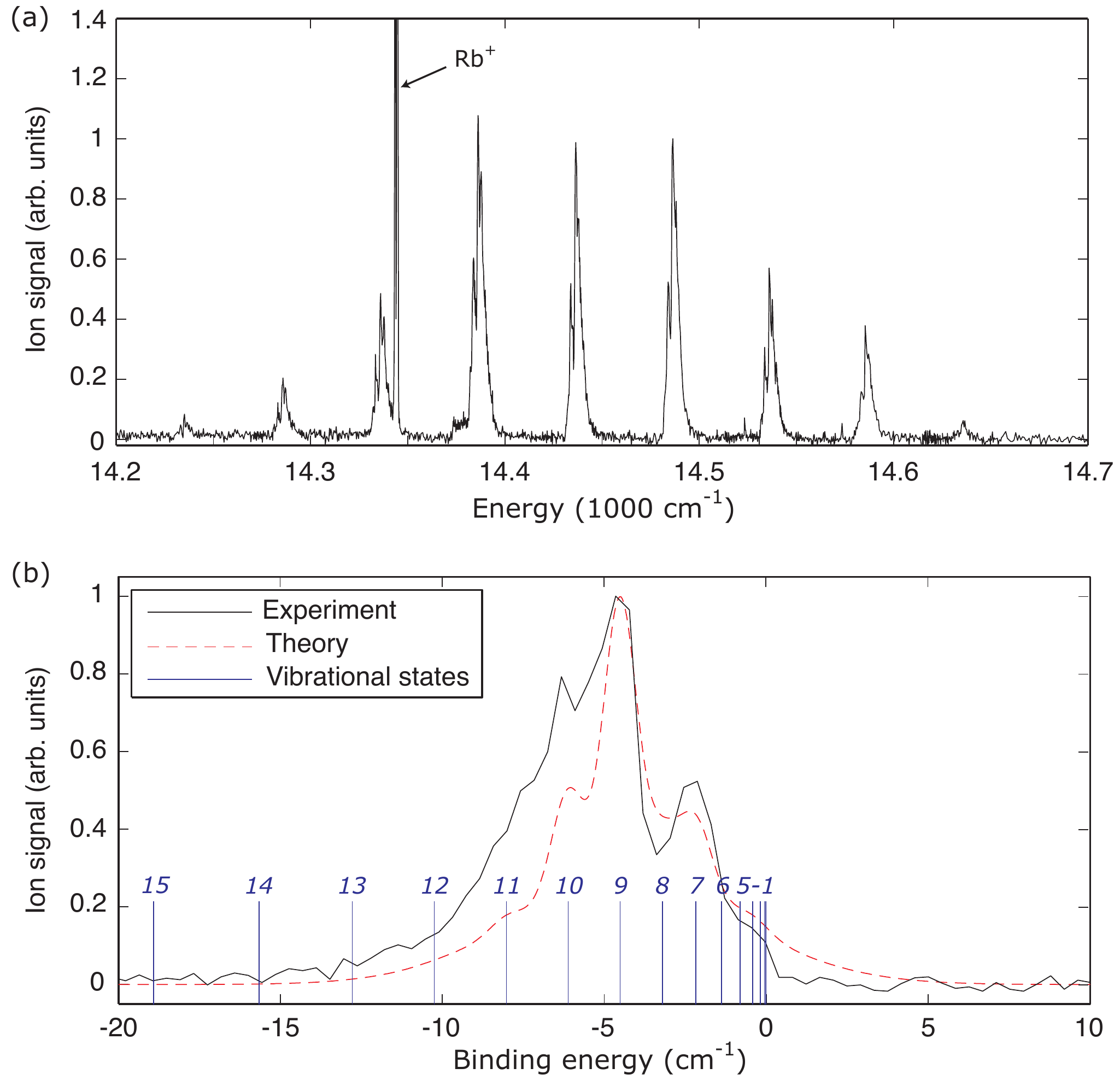}
  \caption{(Colour online). Spectroscopy of the initial state. High-lying vibrational states within the $a^3\Sigma_u^+$ ground triplet formed through MOT laser photoassociation are ionized using a REMPI pathway through the $(2)^3\Sigma_g^+$ potential. The Rb$_2^+$ signal is shown as a function of ionization energy. (a) The full experimental spectrum reveals the large vibrational level spacing within the $(2)^3\Sigma_g^+$ potential. The atomic resonance used for calibration is indicated. The overall spectrum is modulated by the tuning range of the ionization laser. (b) A close-up plot of the spectral feature at \unit[14500]{cm$^{-1}$} reveals the initial population distribution within the $a^3\Sigma_u^+$. A close agreement is found with simulations of the population distribution obtained by MOT photoassociation to the \mbox{5S + 5P$_{3/2}$ $0_{\textrm{g}}^-$} potential. The locations of the highest-lying $a^3\Sigma_u^+$ vibrational levels are shown and enumerated according to increasing binding energy.}
  \label{fig:init-state}
\end{figure}

Figure \ref{fig:init-state}(a) shows the result of a spectroscopic scan of the REMPI ionization signal from the high-lying ground vibrational states populated via background mechanisms. The full spectrum reveals a repeating structure at intervals of \unit[50]{cm$^{-1}$} that corresponds to the spacing of the vibrational levels within the $(2)^3\Sigma_g^+$ excited state (as recorded in Fig.\ 6 of Lozeille et al.\ \cite{Lozeille2006}). The detected Rb$_2^+$ signal strength is modulated by the available ionization laser pulse energy as the wavelength is tuned.

As discussed in Section \ref{sec:REMPI}, the repeated structure within the spectrum reveals the initial population distribution within the $a^3\Sigma_u^+$ ground-state occupancy. A close-up of the spectral feature at \unit[14500]{cm$^{-1}$} is shown in Fig.\ \ref{fig:init-state}(b), together with the calculated population distribution formed via MOT laser photoassociation to the $0_{\textrm{g}}^-$ potential followed by spontaneous decay. The greatest population is found to be contained within the ninth-highest vibrational level, which corresponds to the level with the greatest Franck--Condon overlap from the excited state populated by the MOT light. Around ten bound states are found to be populated in total.

The experimental spectrum is in close agreement with the calculated population distribution formed by MOT laser photoassociation via the \mbox{5S + 5P$_{3/2}$ $0_{\textrm{g}}^-$} potential. This agreement confirms that this, rather than three-body recombination, is the dominant mechanism for background molecule formation within this experimental setup. The detection of a strong background triplet molecule signal implies that this bound initial population, additionally to unbound scattering atom pairs, is a candidate for contribution to the photoassociation dynamics. This is analyzed further in Section \ref{sec:bound-dimers}.

\subsection{Numerical pump-probe simulations}
\label{sec:simulations}

\begin{figure}
  \centering
  \includegraphics [width=\columnwidth]{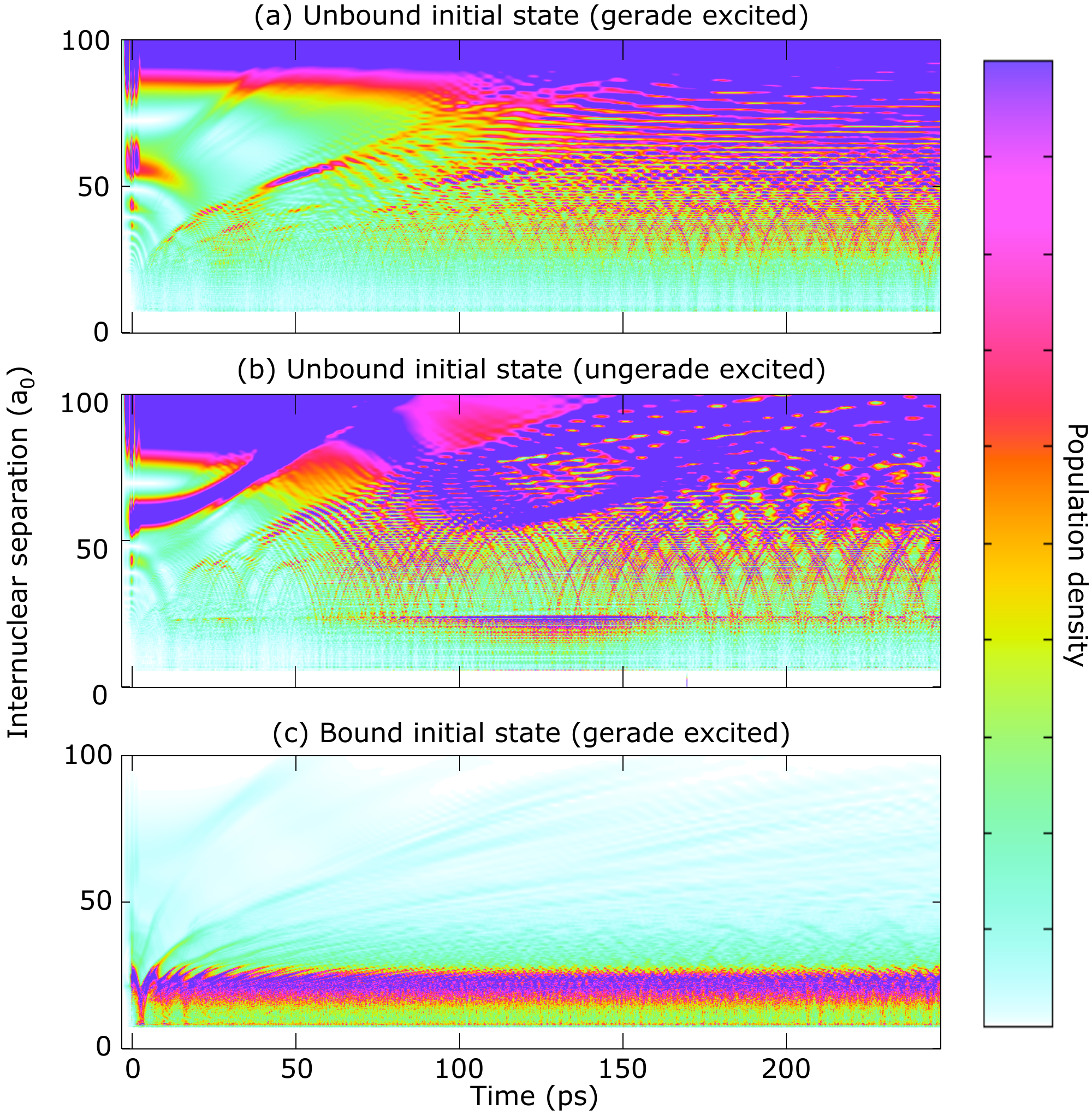}
  \caption{(Colour online). Simulations of the time- and position-dependent excited-state population density induced by the pump pulse. Coherent dynamics occur at a range of internuclear separations (represented by the dark shaded areas) despite the incoherence of the initial states. Results are shown for transitions from an unbound initial population to both gerade (a) and ungerade (b) excited states, as well as from a bound initial population to gerade excited states only (c) (single photon transitions to ungerade states are prohibited from the  $a^3\Sigma_u^+$ potential). }
  \label{fig:dynamics}
\end{figure}

Numerical simulations were conducted of the excited-state dynamics induced by the pump pulse, with consideration given to both the initial-state scenarios discussed above. The time-dependent Schr\"{o}dinger equation was solved for two different initial density matrices: a thermal distribution of scattering states, and the incoherent mixture of bound \mbox{5S + 5P$_{3/2}$ $0_{\textrm{g}}^-$} decay products. The details of the coupled-channel model and the numerical implementation are as discussed in Ref.\ \cite{Martay2009}, but with the pump pulse selected to match Fig.\ \ref{fig:pulses}(b). The time- and position-dependent population density for both initial density matrices is shown in Fig.\ \ref{fig:dynamics}. The dynamics due to transitions from the unbound initial state to excited states of both gerade and ungerade symmetries are shown separately; however dipole selection rules prohibit the population of ungerade excited states from the bound $a^3\Sigma_u^+$ initial state. The excited-state population exhibits a time-varying internuclear separation distribution, and so a position-sensitive measurement should observe a time-varying signal. Close-range, few-picosecond timescale dynamics are predicted for the bound initial state, compared to slower and longer-range dynamics from the unbound initial state.

The time-variation is caused by two effects. Firstly, the population density matrix that describes a time-independent distribution in the ground potential, no longer describes a time-independent distribution in the excited states since different potential energy curves govern its motion. The thermal distribution has a nodal structure that, when transferred to the excited state, forms distinct pockets of population that fall inwards. The anharmonicity of the excited potential at large internuclear separation precludes harmonic motion, but coherent interference effects still cause time-varying oscillations in the population at any internuclear separation. The second effect is that the pump-pulse bandwidth and transition Franck--Condon factors only transfer population in a narrow window in internuclear separation. This further improves the coherence of the excited-state population.

The simulations show that the incoherence of the initial state, whether from thermally distributed scattering states or from preassociated molecules, is not a barrier to the formation of coherently oscillating excited-state population by the pump pulse. The subsequent action of the ionizing probe pulse is less well theoretically understood due to the scope for the ejected electron to carry surplus energy and momentum. Due to the absence of a suitable experimentally verified model for molecular ionization, the probe pulse has been omitted from these calculations. The experimental probe pulse described in Section \ref{sec:ultrafast} is sufficiently short to resolve these dynamics provided that it permits the requisite position-selective measurement.

\subsection{Rubidium pump-probe dynamics}
\label{sec:pump-probe}

\begin{figure}
  \centering
   \includegraphics[width=0.9\columnwidth]{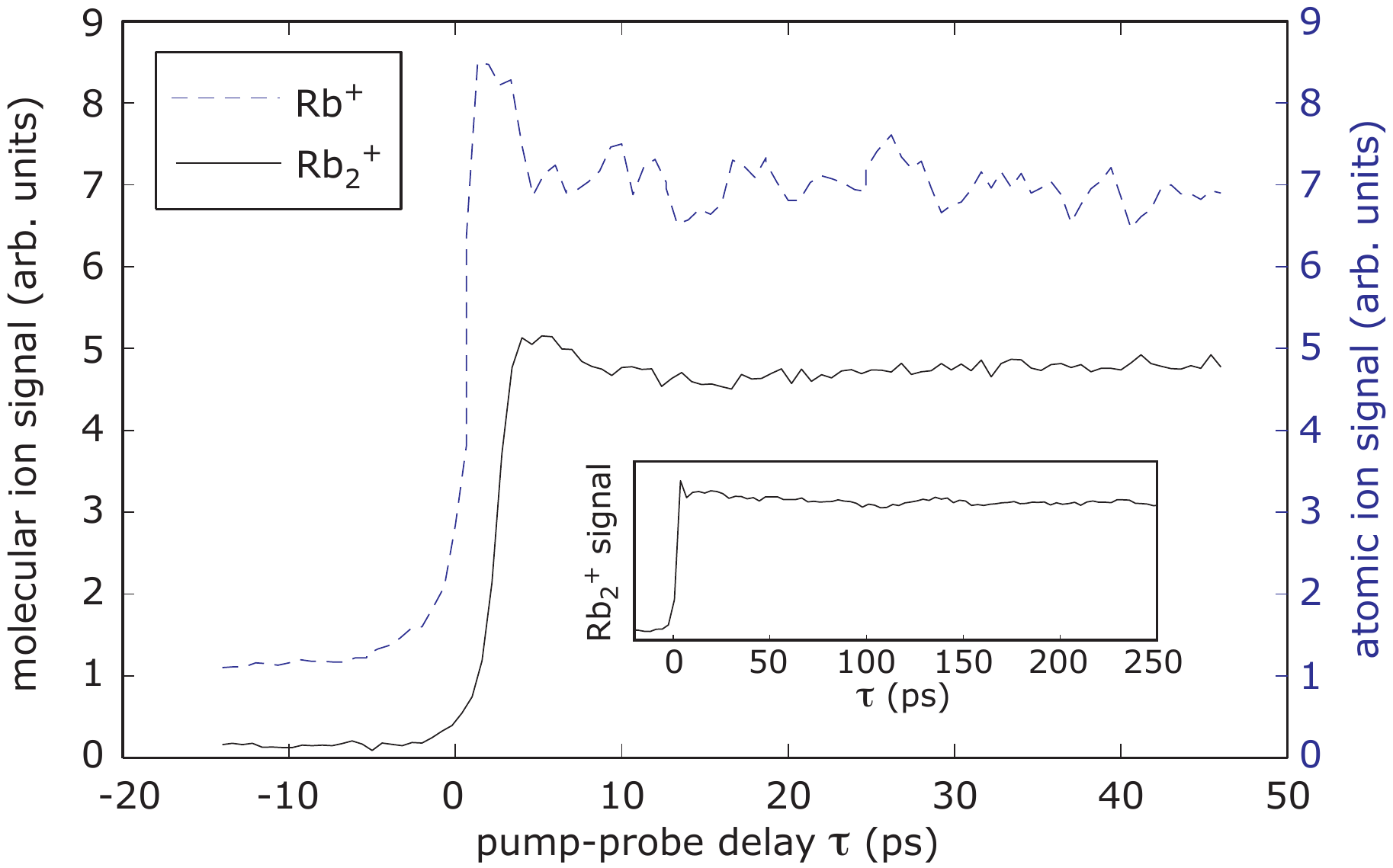}  
  \caption{(Colour online). Typical atomic and molecular ion signals as a function of pump-probe delay (main figure), together with a typical molecular pump-probe signal at larger delays (inset). The atomic ion signal has been scaled for comparison. Both atomic and molecular signals exhibit a step at zero delay and a peak at small positive delays due to transient effects induced by the pump pulse. Fourier analysis of the subsequent signal reveals no dominant oscillatory timescales.}
  \label{fig:pu-pr}
\end{figure}

Figure \ref{fig:pu-pr} shows typical atomic and molecular pump-probe signals as a function of pump-probe delay. A low pump-pulse energy of \unit[85]{nJ} was used in order to allow atomic and molecular dynamics to be monitored simultaneously without excessive atomic ionization. The inset also shows a typical molecular pump-probe signal at longer delays for the higher pulse energy of \unit[400]{nJ}. The absolute Rb$^+$ and Rb$_2^+$ signal scalings employed in the plots are arbitrary and have been chosen to take a similar magnitude. Both the Rb$^+$ and Rb$_2^+$ experimental signals show a distinct step at zero delay: the signals are greater for a pump-probe sequential pulse order, though significant off-resonant atomic ionization is still evident with probe-pump timings (Fig.\ \ref{fig:pu-pr}(a)). In the Rb$_2$ pump-probe experiment of Salzmann et al.\ \cite{Salzmann2008}, the sharp spectral cut applied to the pump pulse introduced a long tail into the pump-pulse envelope in the corresponding temporal domain. In exciting an atom pair, the pump pulse induces an electronic dipole which interacts coherently with this long temporal tail, causing oscillations in the excited-state population and hence the Rb$_2^+$ signal. This behaviour is similar to the observations of Zamith et al.\ in a Rb vapour cell \cite{Zamith2001}. These dipole transient effects are not apparent in the long-timescale pump-probe signals of this work due to the employment of a less sharp spectral cut of the pump pulse.

%Figure \ref{fig:pu-pr}(a) shows typical atomic and molecular pump-probe signals as a function of pump-probe delay. A low pump-pulse energy of \unit[85]{nJ} was used in order to allow atomic and molecular dynamics to be monitored simultaneously without excessive atomic ionization. Fig.\ \ref{fig:pu-pr}(b) shows a typical molecular pump-probe signal obtained with the higher  pump-pulse energy of \unit[400]{nJ}, together with simulations of the molecular dynamics for two initial state cases: the bound triplet population characterized in Section \ref{sec:initial-state} and an unbound scattering state. The absolute Rb$^+$ and Rb$_2^+$ signal scalings employed in the plots are arbirary and have been chosen to take a similar magnitude (in the case of Fig.\ \ref{fig:pu-pr}(a)) and to fit to the theoretical simulations (Fig.\ \ref{fig:pu-pr}(b)).

Fourier analysis of the positive-delay signal that follows the coherent transient peak does not reveal any dominant characteristic oscillatory periods. This is in contrast with the picosecond-timescale coherent excited-state dynamics predicted by the simulations (which the \unit[400]{fs} probe pulse was short enough to resolve), particularly for the case of the bound initial state. The calculations of Section \ref{sec:simulations} revealed a wide range of conditions under which coherent oscillations may be formed. It therefore seems likely that the absence of such oscillations in the pump-probe signal can be attributed to an insufficiently position-dependent measurement being effected by the probe, for which there are two likely causes. Firstly, the ionization mechanism of the probe pulse is not as well understood as for neutral-neutral transitions (as outlined in Section \ref{sec:simulations}). In order to attain a sufficiently short pulse to resolve the dynamics, a large bandwidth is required which may therefore compromise selectivity with respect to internuclear separation. Other indirect mechanisms, such as probe-pulse excitation to neutral Rydberg states that couple to the Rb$_2^+$ potential, may also complicate the process. Secondly, the detection mechanism may not be sensitive to very long-range excited-state population. Further information is thus required concerning the nature of the initial population addressed by the pump pulse. These issues are further discussed below in Section \ref{sec:bound-dimers}.

\subsection{Bound excited-state dimers}
\label{sec:bound-dimers}

\begin{figure}
  \centering
   \includegraphics[width=0.9\columnwidth]{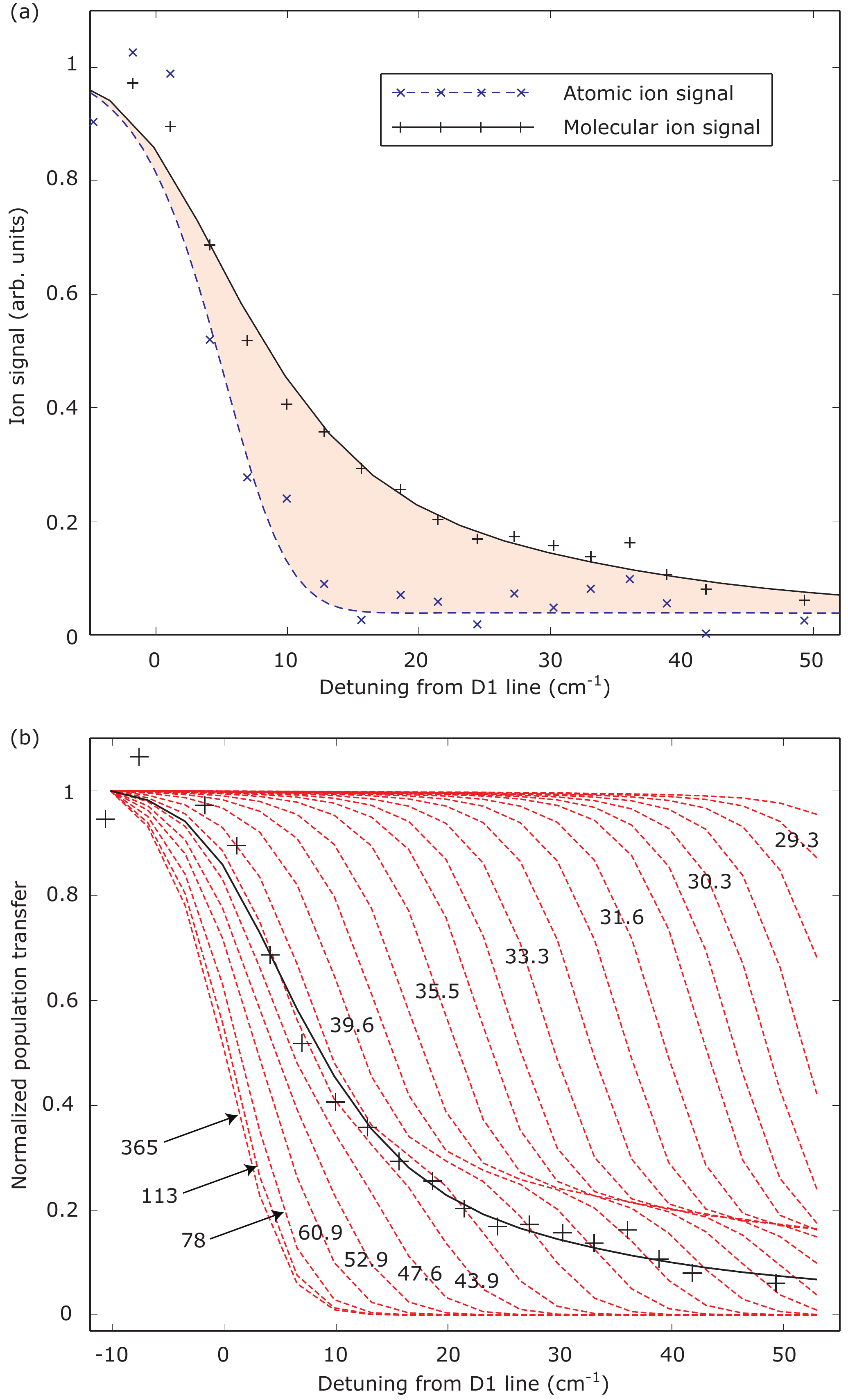}
  \caption{(Colour online). (a) Normalized atomic and molecular signals measured at a fixed pump-probe delay for different detunings of the pump-pulse spectral cut from the D1 line. The shaded area indicates evidence for the formation of bound excited-state dimers. (b) Normalized theoretical calculations of the excited-state population transfer induced by a pump pulse as a function of spectral cut detuning (red dashed lines). The different series represent the behaviour of pairs of atoms of different initial separations within the $a^3\Sigma_u^+$ ground state (labelled in atomic units). The experimental molecular ion signal from (a) is plotted together with an optimized theoretical fit for a distribution of initial separations around \unit[50]{a$_0$}.}
  \label{fig:bound_mols}
\end{figure}

The Rb$_2^+$ ion signal did not appreciably decay over the delays of hundreds of picoseconds that these experiments considered. In order to demonstrate that this large-delay signal is attributable to bound molecular population within the excited-state manifold (rather than further coherent transient effects or photoassociation of an unbound 5S and 5P atom to bound molecular ions), a further study was performed. A fixed, positive pump-probe delay of \unit[40]{ps} was selected and the atomic and molecular ion signals were recorded as a function of the position of the pump-pulse spectral cut.

As the pump-pulse spectral cut was further detuned from the \mbox{5S + 5P$_{1/2}$} atomic asymptote (increasing the binding energy of the energetically accessible excited states), the Rb$^+$ signal was observed to fall off more quickly than the Rb$_2^+$ signal (\mbox{Fig.\ \ref{fig:bound_mols}(a)}). Indeed, atomic ion formation became negligible at detunings beyond \unit[10]{cm$^{-1}$}, while a substantial number of molecular ions continued to be formed. The fast decline in Rb$^+$ signal is caused by the loss of spectral intensity resonant with the atomic transition. Hence less atomic population is excited and subsequently ionized. The Rb$_2^+$ signal persists, however, due to excitation into bound vibrational levels below the atomic asymptote. The difference between the Rb$^+$ and Rb$_2^+$ data series (shaded) therefore provides strong evidence for the formation of bound molecular wavepackets by the pump pulse.

In order to study the initial state that contributed to this pump-probe signal, theoretical calculations were performed of the excited-state population transfer induced by a realistic experimental pump pulse as a function of spectral cut detuning (\mbox{Fig.\ \ref{fig:bound_mols}(b)}), and repeated for pairs of atoms with a range of initial separations. The asymptotic long-range form of the excited-state potentials scales as $R^{-3}$ due to dipole-dipole interactions, whereas the ground-state potentials scale as $R^{-6}$ due to van der Waals interactions. The pump-pulse transition energy therefore becomes more red-detuned from the atomic asymptote at closer range. As a consequence, these simulated population transfers fall off more slowly with spectral cut detuning for closer range initial populations.

\begin{figure}
  \centering
  \includegraphics[width=0.9\columnwidth]{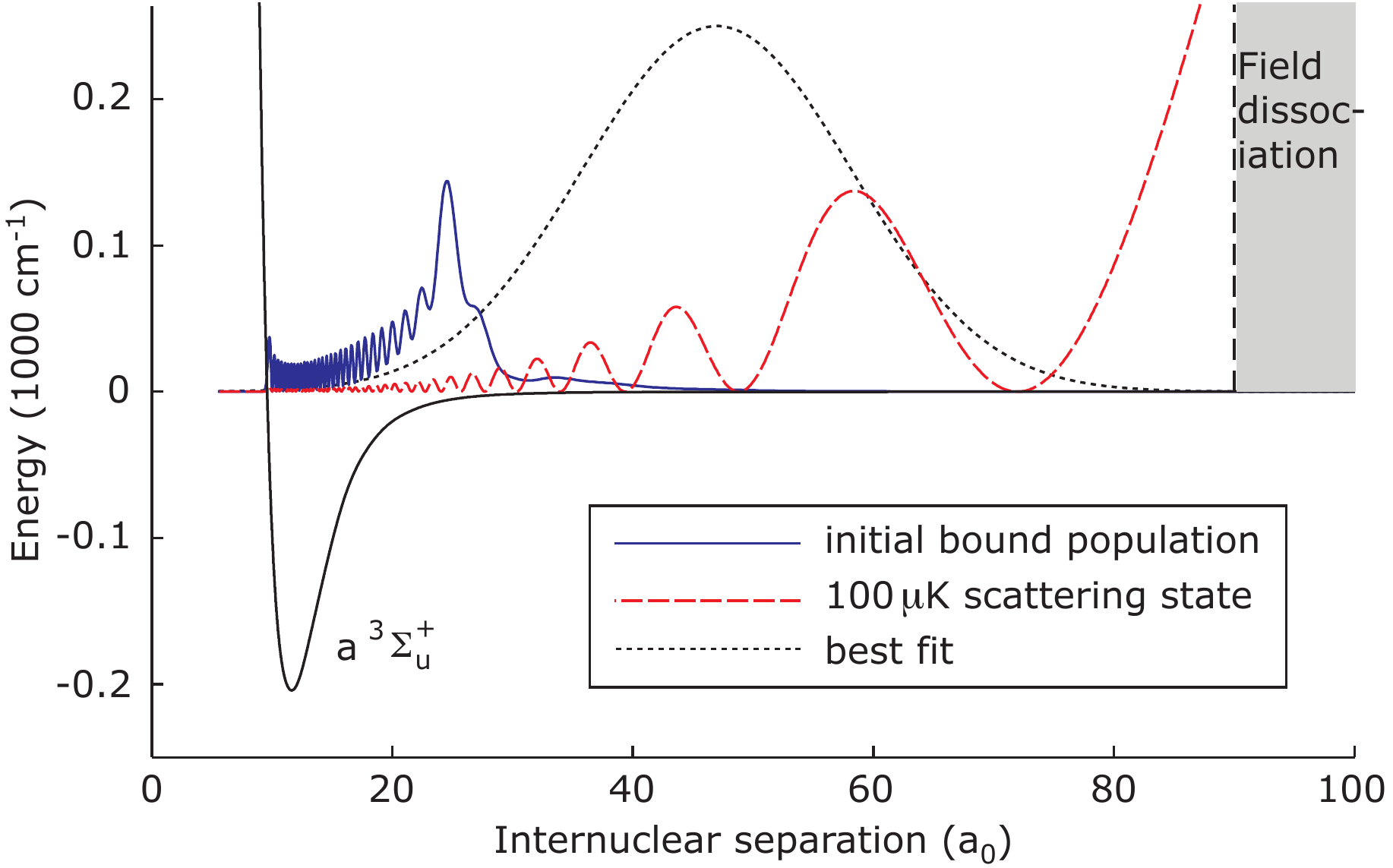}
  \caption{(Colour online). The population distributions of the bound background molecules characterized in Section \ref{sec:initial-state} and a \unit[100]{$\mu$K} scattering state, together with an analytical best fit of the population that contributes to the pump-probe signal. The contributing population is of longer range than the background molecules, but is consistent with the association of unbound scattering atoms by the pump pulse. Dissociation of molecular ions by the TOF electric field is calculated to account for the lack of contribution of very long-range scattering atom pairs to the Rb$_2^+$ signal.}
  \label{Pu-pr_init}
\end{figure}

The atom pairs contributing to this pump-probe process will realistically be distributed over a range of internuclear separations rather than located at a single point; indeed, the fact that the experimental data span a number of these theoretical series corroborates this picture. Excellent agreement was found when the initial population distribution was fitted to the experimental data (\mbox{Fig.\ \ref{fig:bound_mols}(b)}). Several analytical forms of the initial distributions were found to fit the data closely; all shared the common characteristic that population was concentrated at internuclear separations of \unit[30-60]{a$_0$} (see \mbox{Fig.\ \ref{Pu-pr_init}}). By contrast, the outer turning point of the most populated triplet vibrational state according to the study of Section \ref{sec:initial-state} is located at \unit[25]{a$_0$}. This provides evidence that the population addressed by the pump and probe pulses is different from the background triplet distribution photoassociated by the MOT light. Instead, the inferred distribution is consistent with the association of unbound atoms, coinciding with several antinodes of the scattering wavefunction. In this case, the detuning of the probe pulse from the Rb$^+$ asymptote might account for the lack of a contribution from the very long-range portion of the scattering wavefunction. As an alternative explanation, our calculations indicate that the TOF electric field is sufficient to dissociate molecular ions (causing \mbox{$\textrm{Rb}_2^+ \rightarrow \textrm{Rb} ^+ + \textrm{Rb}$}) of longer range than around \unit[90]{a$_0$}.

\section{Discussion}
\label{sec:discussion}

We have performed a spectroscopic measurement of the initial state of the preformed molecules within our ultracold $^{85}$Rb$_2$ photoassociation experiments. The twin mechanisms of three-body recombination and trapping light photoassociation are known to contribute to a background population of loosely bound ground-state triplets within the MOT with relative impacts that are determined by the particular experimental conditions. Our measurement of the initial ground triplet-state occupancy reveals a distribution that closely matches the photoassociative mechanism, confirming this to be the dominant contribution in this case.

We have carried out a pump-probe study of the excited-state dynamics of Rb$_2$ dimers as the first step towards a fully coherent pump-dump transfer to the ground state. An ultrafast pump pulse was spectrally filtered to populate bound excited states only, and the resultant wavepacket was allowed to evolve for a controllable delay before being ionized with a probe pulse. By observing the large-delay behaviour of the pump-probe signal after the influence of initial transient effects, we have shown the formation of excited-state population that is demonstrably bound in character.

We have characterized the population that contributes to the pump-probe signal and have found that its behaviour with respect to spectral cut location is consistent with the association of unbound scattering atom pairs, rather than the excitation of background triplet molecules. Calculations indicated that the bulk of the excited-state dynamics induced from such an unbound initial state would occur at larger internuclear separations than are resonantly addressed by the probe pulse. This may therefore be a factor in the lack of experimental observation of pump-probe dynamics.

The ultimate goal, however, remains the coherent control of this population so as to maximize the Franck-Condon overlap to the dump-pulse target state after a suitably chosen delay; a signal presenting evidence of wavepacket oscillations is therefore required. Simulations reveal that pre-associated molecules in an incoherent mixture of states can be made to oscillate coherently with \unit[5]{ps} timescales at relatively close range using a single ultrafast pump pulse \cite{Martay2009}. By contrast, the classical oscillation period of molecules photoassociated from scattering atom pairs is greater by an order of magnitude or more, and occur predominantly at long range where detection is particularly difficult. %It represents a far greater challenge to design a probe pulse capable of selectively detecting these long-range dynamics, as the required energies are much closer to the atomic resonances, and change much more slowly with internuclear separation than at close range.
Increasing the signal-to-noise size at the larger pump-pulse detunings that address the relatively close-range pre-formed molecules should therefore allow this coherent behaviour to be revealed and controlled.

Since the photoassociation rate scales in proportion to the product of the density and the number of trapped atoms, we identify these two parameters as targets for an improvement to our experimental apparatus.  \mbox{Figure \ref{fig:bound_mols}(b)} indicates that an order-of-magnitude improvement in photoassociation rate would provide a sufficient signal-to-noise ratio to probe the background molecular population located around \unit[30]{a$_0$}. To this end we have trialled a dark spontaneous-force optical trap (SPOT) MOT \cite{Ketterle1993} in order to improve density. This protocol is difficult to implement effectively in our existing Rb MOT, however, as the low branching ratio in Rb to the dark state requires high extinction of the repumper, and we consequently saw no improvement in signal. Instead, we are constructing a new MOT apparatus with which we expect to obtain this necessary improvement in atom number and density.

\section{Conclusion}
\label{sec:conclusions}

There is much current research interest devoted to the preparation of translationally ultracold molecules in their absolute ground electronic and rovibrational states. A broad range of strategies is at the disposal of the experimentalist in the pursuit of this goal. Some significant recent advances have used the approach of associating ultracold atoms, with the experimental realization of $v=0$ dimers in their absolute ground electronic states via techniques such as STIRAP or incoherent optical cycling. These breakthroughs for certain specific species are complemented by coherent control strategies, which are expected to be broadly applicable to a range of systems without the need for fortuitous initial bound-state preparation pathways, a detailed knowledge of the system spectroscopy or incoherent transfer processes.

In this work, we have demonstrated the formation of bound excited-state dimers by an ultrafast photoassociation pulse. This represents an important milestone in the design of a coherent pump-dump transfer strategy. We have determined the initial state addressed by the pump pulse and demonstrated that it predominantly comprises longer-range initial population than those background molecules pre-associated by the trapping light. We have identified improvements in MOT number and density as being critical to allow the detection and control of the coherent excited-state dynamics of closer-range population.

\begin{acknowledgments}
The authors are grateful to Jordi Mur-Petit for valuable discussions and suggestions. \mbox{H.\ E.\ L.\ M.\ }acknowledges help from Thorsten K\"{o}hler. This work was supported by EPSRC grant number EP/D002842/1, Marie Curie Initial Training Network CA-ITN-214962-FASTQUAST, and Bulgarian NSF grant number WU-301/07.
\end{acknowledgments}

\end{document}